\documentclass[twocolumn,english,journal]{IEEEtran}
\ifCLASSINFOpdf
\else
\fi

\usepackage[T1]{fontenc}
\usepackage{babel}
\usepackage{array}
\usepackage{calc}
\usepackage{amsmath}
\usepackage{amsthm}
\usepackage{amssymb}
\usepackage{graphicx}
\usepackage{cite}
\usepackage{algorithm}
\usepackage{subfigure}
\usepackage{color}

\usepackage{enumitem}
\usepackage{multirow}
\usepackage{hhline}
\usepackage{float}
\usepackage[justification=centering]{caption}

\usepackage[unicode=true,
bookmarks=true,bookmarksnumbered=true,bookmarksopen=true,bookmarksopenlevel=1,
breaklinks=false,pdfborder={0 0 0},backref=false,colorlinks=false]
{hyperref}
\hypersetup{pdftitle={Your Title},
	pdfauthor={Your Name},
	pdfpagelayout=OneColumn, pdfnewwindow=true, pdfstartview=XYZ, plainpages=false}

\makeatletter

\providecommand{\tabularnewline}{\\}

\let\oldforeign@language\foreign@language
\DeclareRobustCommand{\foreign@language}[1]{%
	\lowercase{\oldforeign@language{#1}}}

\makeatother

\allowdisplaybreaks

\makeatletter
\newcommand{\thickhline}{%
	\noalign {\ifnum 0=`}\fi \hrule height 1.5pt
	\futurelet \reserved@a \@xhline
}
\newcolumntype{"}{@{\hskip\tabcolsep\vrule width 1pt\hskip\tabcolsep}}
\makeatother

\newcolumntype{P}[1]{>{\centering\arraybackslash}p{#1}}
\newcolumntype{M}[1]{>{\centering\arraybackslash}m{#1}}

\begin{document}
	%
	\title{Vulnerability Assessment of $N-1$ Reliable Power Systems to False Data Injection Attacks}

	
	
	%
	\author{\IEEEauthorblockN{Zhigang Chu,
			Jiazi Zhang, 
			Oliver Kosut, and
			Lalitha Sankar\\}
		\IEEEauthorblockA{School of Electrical, Computer and Energy Engineering\\
			Arizona State University}}


	\maketitle
	\pagestyle{plain}
	\begin{abstract}
		This paper studies the vulnerability of large-scale power systems to false data injection (FDI) attacks through their physical consequences. Prior work has shown that an attacker-defender bi-level linear program (ADBLP) can be used to determine the worst-case consequences of FDI attacks aiming to maximize the physical power flow on a target line. Understanding the consequences of these attacks requires consideration of power system operations commonly used in practice, specifically real-time contingency analysis (RTCA) and security constrained economic dispatch (SCED). An ADBLP is formulated with detailed assumptions on attacker's knowledge, and a modified Benders' decomposition algorithm is introduced to solve such an ADBLP. The vulnerability analysis results presented for the synthetic Texas system with 2000 buses show that intelligent FDI attacks can cause post-contingency overflows.
	\end{abstract}
	

	%
	\IEEEpeerreviewmaketitle
	\global\long\def\figurename{Fig.}
	\global\long\def\tablename{TABLE}
	
	\vspace{-0.4cm}
	\section{Introduction}
    The efficiency and intelligence of modern electric power systems are increasing rapidly with integration of real-time monitoring, sensing, communication and data processing. This integration is accomplished via a cyber layer consisting of the supervisory control and data acquisition (SCADA) system in conjunction with the energy management system (EMS). SCADA monitors the physical system, collects measurements, and sends them to the control center. In the EMS, state estimation (SE) estimates the voltage magnitudes and angles from measurements. This estimate along with the subsequent data processing, optimization and communication, specifically real-time contingency analysis (RTCA) \cite{Mittal2011} and security constrained economic dispatch (SCED) \cite{SCED2006}, allow for real-time control of the power systems. 
	
	However, the integration of the cyber layer also increases the threat of cyber-attacks on power systems that could lead to severe physical consequences, as illustrated by the recent cyber-attack in Ukraine (see \cite{UkraineAttack}). Therefore, it is crucial to develop techniques to detect and thwart potential attacks, which requires evaluating system vulnerability to credible attacks. Assessing consequences of possible attacks is extremely instructive for system operators, and is important for secure power system operations. 
	
	This paper focuses on false data injection (FDI) attacks, wherein a malicious attacker replaces a subset of SCADA measurements (power flows and injections) with counterfeits. 
	FDI attacks can be designed to target system states \cite{Liu2009,Kosut2011,Hug2012}, system topology \cite{Kim2013a,Jzhang2016}, and energy markets \cite{Moslemi2018}.  Optimization problems have been proposed to design FDI attacks that aim to maximize line power flow \cite{Liang2015}, change locational marginal prices \cite{Jia2014}, or maximize operating cost \cite{Yuan11}. However, the results have only been demonstrated for small systems, and do not include the effects of $N-1$ reliability constraints. $N-1$ reliable system operations typically involve RTCA to generate security constraints, and SCED to re-dispatch the generators with the security constraints in the most economic sense. In this paper, we focus on the worst-case FDI attacks that maximize power flow on a target line, but our goal is to evaluate vulnerability of significantly larger systems (\emph{i.e.} thousands of buses) with AC SE, RTCA, and SCED.
	
	The authors of \cite{Liang2015} introduce an unobservable FDI attack that re-distributes the loads by changing SCADA measurements, to trigger generation re-dispatches that result in physical overflow on a target line. The authors show that the worst-case attack can be found using an attacker-defender bi-level linear program (ADBLP), wherein the first level models the attacker's objective and limitations, and the second level models the system response through DC optimal power flow (OPF). In this paper, we formulate a similar ADBLP with the second level modeled as SCED instead of OPF, taking into account the security constraints generated by RTCA. We also discuss in detail the required knowledge of an attacker to design such an unobservable attack.
	
	Techniques to solve ADBLPs with applications to power systems have been studied in \cite{Yuan2012,Alderson2011}, but are limited to scenarios with the same objective for both levels, and hence, their techniques cannot be applied to either the problem in \cite{Liang2015} or its generalization considered here for large power systems. An ADBLP can be reformulated as a mathematical program with equilibrium constraints (MPEC) \cite{Mathiesen1985} by replacing the second level by its Karush-Kuhn-Tucker (KKT) conditions. However, MPECs are non-convex and hard to solve efficiently in general \cite{Luo1996}. The MPEC from the ADBLP can be further reformulated as a mixed-integer linear program (MILP) by rewriting the complementary slackness constraints as mixed-integer constraints. As the system size increases, this MILP becomes harder to solve due to the increasing number of binary variables. In this paper, we introduce an efficient algorithm that utilized Benders' decomposition to solve ADBLPs.

	The contributions of this paper are as follows:  
	\begin{enumerate}
		\item Knowledge requirement for attacker to design unobservable attacks in the presence of RTCA and SCED;
		\item Attack design ADBLP modeling SCED as the second level problem;
		\item Modified Benders' decomposition algorithm to solve ADBLPs;
		\item Simulations of the designed attacks on the synthetic Texas system to evaluate the consequences of such attacks.
	\end{enumerate}
	
	The remainder of this paper is organized as follows. Sec. \ref{sec: SE and models} describes the power system measurement model and unobservable attack model. Sec. \ref{sec:AtkAssumption} discusses attacker's requirement to design and launch worst-case unobservable attacks. Sec. \ref{sec:ADBLP} introduces an ADBLP to find worst-case attacks. Sec.\ref{sec:Benders} describes the modified Benders' decomposition algorithm to solve any ADBLP. Sec. \ref{sec:Simulation} presents the simulation results on the Texas system. Sec. \ref{sec:conclusion} concludes the paper and considers directions of future work.
	
	\section{\label{sec: SE and models}System and Attack Model}
	\subsection{EMS Operation\label{sec:EMS_OP}}
\begin{figure}[h]
		\centering{}\includegraphics[trim=0 0.2cm 0 0.2cm, scale=0.45]{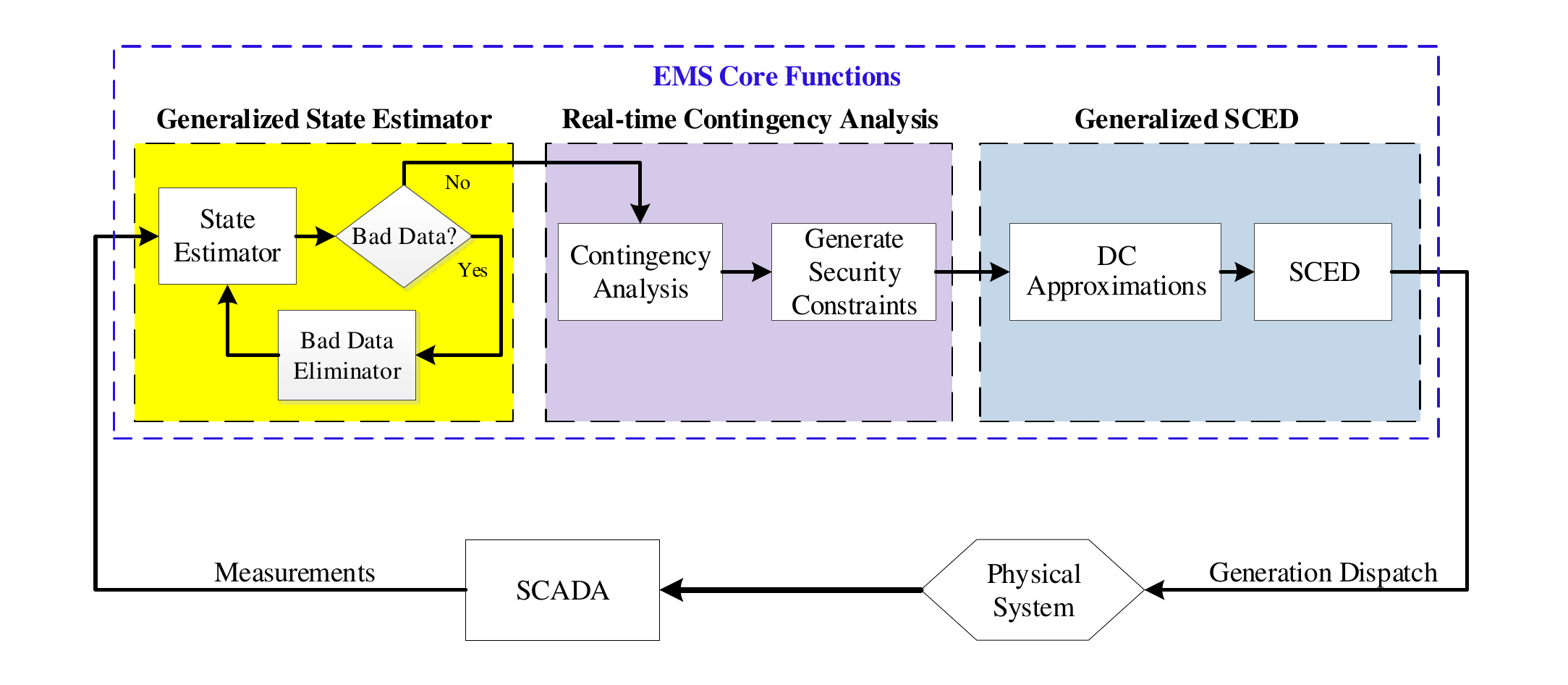}\protect\protect\caption{EMS operation with SE, RTCA, and SCED. \label{fig:EMS_op}}
		\vspace{-0.5cm}
	\end{figure}

	In this paper, we consider an EMS with three core functions operating in the order of SE, RTCA, and SCED. The EMS operating structure is illustrated in Fig. \ref{fig:EMS_op}. Power system measurement data collected by SCADA are sent to SE, which estimates the complex voltages after eliminating noise and bad measurements. Given the generator set points, SE also estimates the load values in the system. Modern power systems typically require $N-1$ reliability, \textit{i.e.,} the system must operate with no violations if a contingency occurs (one of the system components, generators or branches, is out of service). RTCA simulates one power flow under each contingency $k$. We say a branch has a \textit{warning} if its power flow is above a threshold $\tau$ but less than its limit, while a branch has a \textit{violation} if its power flow exceeds its limit. Note that in base case, the limit is the long-term line rating, while in contingency case it is the short-term rating. Each warning and violation generates one line limit constraint to be modeled in SCED. In contingency cases, these constraints are called security constraints. SCED takes all these constraints and solves an optimization problem to determine the most economic generation dispatch that ensures $N-1$ reliable operation. 
	\subsection{Measurement Model}
	We model the power system with $n_b$ buses, $n_g$ generators, and $n_m$ measurements. 
	The SCADA system measurement model is given by
	\begin{equation}
		z=h(x)+e\label{eq:DCMeasurement}
	\end{equation}
	where $z$ is the $n_{m}\times1$ measurement vector; $x$ is the $2n_b\times1$ vector of bus voltage magnitudes and angles (states); $h(\cdot)$ is the non-linear relationship between measurements and states; $e$ is the $n_{m}\times1$ vector of measurement noise, whose entries are assumed to be jointly distributed as $\mathcal{N}$(0,$R$) where $R=\text{diag}(\sigma_{1}^{2},\sigma_{2}^{2},\ldots,\sigma_{n_{m}}^{2})$.
\vspace{-0.2cm}
	\subsection{\label{AttackModel}Unobservable Attack Model}
	An $n_{m}\times 1$ measurement attack vector $a=\bar{z}-z$ is defined to be \emph{unobservable} to the residual-based bad data detector (BDD) if $a=h(x+c)-z$, where $\bar{z}$ is the vector of the false measurements created by the attacker and $c$ is the state attack vector \cite{Hug2012}. Given $c$, an attack subgraph $\mathcal{S}$ can be constructed as in \cite{Hug2012}, such that the non-zero entries of $a$ are all within $\mathcal{S}$.
%
%
	By modifying measurements only in $\mathcal{S}$, the attacker can arbitrarily spoof the states of center buses (load buses corresponding to non-zero entries of $c$) without detection. The attack causes the system estimated load to re-distribute between load buses within $\mathcal{S}$, while the total load remain unchanged.

    \vspace{-0.1cm}
	\section{\label{sec:AtkAssumption}Attacker Assumptions}
	Among all unobservable FDI attacks, the most dangerous ones are those with serious physical consequences. In this paper, we focus on a class of unobservable attacks where the attacker maliciously changes the SCADA measurements to maximize the power flow on a target line, and possibly cause overflow. The authors of \cite{Liang2015} introduce an ADBLP to determine the worst-case unobservable line overflow attack, wherein the first level models the attacker's objective and limitations, while the second level models the system response via DCOPF. Assuming the attacker has knowledge of (i) the complete network topology (including line parameters and ratings) and load information, and (ii) the cost, capacity, and operational status of all generators in the system, the authors show that unobservable attacks found using this optimization successfully result in generation re-dispatches that cause line overflows on the IEEE RTS 24-bus system. 
	
	However, modern power systems typically do not use DCOPF to re-dispatch the generation, but rather operates as outlined in Sec. \ref{sec:EMS_OP}. An attacker who gains knowledge of EMS operations has an advantage to accurately predict the system response. In other words, if the attacker is able to perform the same RTCA and SCED as the system does, it can design attacks that maximize the consequences. This is a stronger assumption than that in \cite{Liang2015}, because in addition to having access to the database of the control center, now the attacker further knows the algorithms and assumptions used by the system. While this is a stronger requirement, it is valuable to understand how the system is resilient against such strong adversaries through this worst-case approach.

	In RTCA, the attacker needs to know the power flow algorithm used to get the same post-contingency flows on all lines, as well as the threshold $\tau$ as described in Sec. \ref{sec: SE and models}, to determine the security constraints to be included in SCED. In SCED, the attacker should know how the system models the constraints, as different system operators may implement SCED differently. We assume the attacker has full knowledge of RTCA and SCED implementation in the EMS, in particular:
	\begin{enumerate}
		\item Contingency ratings of the branches;
		\item Loss handling method;
		\item Ramp rates and reserve costs of all generators;
		\item Reserve policy and requirements;
		\item Criteria to determine which base case line limits are to be modeled. This can be the same threshold as $\tau$ in post-contingency case, but can also be different; 
		\item Branch flow calculation method in both base case and contingency case;
		\item Load shedding policy and costs.
	\end{enumerate}
	While it seems unrealistic to gain such knowledge, it is not entirely impossible, since such complex systems involve sophisticated (even nation-state) attackers that can exploit or have access to insider knowledge \cite{UkraineAttack,StuxnetIran}. Again, this is the worst-case assumptions, and therefore, resilience of the system to such worst-case attacks can serve as an upper bound on risks to the system operations.


	\section{\label{sec:ADBLP}ADBLP to Find Worst-case Attack}
	In this section, we introduce an ADBLP similar to that in \cite{Liang2015} to find worst-case line overflow attacks. The first level models the attacker's objective and limitations, while the second level models the system response via SCED. We focus on RTCA that simulates branch contingencies (excluding radial branches), and reports corresponding security constraints to SCED. Contingency $k$ indicates that branch $k$ is out of service. The attacker is assumed to be able to perform RTCA and pick a target line $l$ to maximize its power flow when target contingency $k_t$ occurs, and possibly create overflow. Without loss of generality, we assume the flow on $l$ is positive; if it is not the case, its absolute value can be maximized. In the formulation below we assume the attacker aims to maximize post-contingency power flow on the target line, but the base case power flow can also be maximized. Since SCED is DC, the voltage magnitudes are all considered to be 1 p.u., and hence, $c$ is an $n_b\times 1$ attack vector on the voltage angles. \ref{sec:AtkAssumption}.
	
	The ADBLP takes the following form:
	\begin{subequations} \label{eq:ADBLP}
	\begin{flalign}
		\hspace{-0.3cm}\underset{c}{\text{maximize}}\: \hspace{0.2cm} & P_{l,k_t}-\sigma\left\Vert c\right\Vert _{1}\label{eq:Obj1_MaxPF}\\
		\notag \text{subject to}\hspace{0.2cm}\;\\
		& \hspace{-1cm}P_{l,k_t}=\text{OTDF}_{k_t}^l(G_{B}P_{G}^{*}-P_{D}) \label{eq:Physical_PF}\\
		& \hspace{-1cm}\|c\|_1 \le N_1 \label{eq:con_L1_norm}\\
		& \hspace{-1cm}-L_{S} P_{D}\le Hc\le L_{S} P_{D}\label{eq:con_loadshift}\\
		& \hspace{-1cm}\left\{P_{G}^{*}\right\} =\text{arg}\left\{ \underset{P_{G},R_G,P,P_k}{\text{min}}\: C_{G}\left(P_{G}\right)+C_RR_G\right\} \label{eq:OBJ_MINCOST}\\	
		&\notag \hspace{-0.9cm} \text{subject to}\\	
		& \hspace{-0.3cm}\begin{array}{lr} 
		\sum_{g=1}^{n_{g}}P_{Gg}=\sum_{i=1}^{n_{b}}P_{Di}\end{array}\label{eq:con_nodebalance}\\
		& \hspace{-0.2cm}\bar{P}=P_0+\text{PTDF}(G_B(P_G-P_{G0})+Hc)\label{eq:con_basePF}\\
		& \hspace{-0.2cm}\bar{P}_k=P_{k0}+\text{OTDF}_k(G_B(P_G-P_{G0})+Hc)\label{eq:con_ctgcPF}\\
		& \notag \hspace{0.6cm} + \text{LODF}_k\cdot\text{PTDF}^k\cdot Hc, \forall k\\
		& \hspace{-0.2cm}-P_\text{max}\le \bar{P}\le P_\text{max} \label{eq:con_basePmax}\\
		& \hspace{-0.2cm}-P_{k,\text{max}}\le \bar{P}_k\le P_{k,\text{max}}, \forall k \label{eq:con_ctgcPmax}\\
		& \hspace{-0.2cm} P_G \ge \text{max}\{P_{G0}-M_GT_h, P_{G,\text{min}}\}\label{eq:con_PGmin}\\
		& \hspace{-0.2cm} P_G \le \text{max}\{P_{G0}+M_GT_h, P_{G,\text{max}}\}\label{eq:con_PGmax}\\
		& \hspace{-0.2cm} 0 \le R_G \le M_GT_r\label{eq:con_reserve}\\
		& \hspace{-0.2cm} P_G+R_G\le P_{G,{\text{max}}}\label{eq:con_genlimit}\\
		& \hspace{-0.2cm} \begin{array}{lr} 
		\sum_{g=1}^{n_{g}}R_{Gg}\ge P_{Gg}+R_{Gg}\end{array}, \forall g\label{eq:con_genctgc}
	\end{flalign}
	\end{subequations}
	where the variables are:
	\begin{description}[leftmargin=1.8cm,style=multiline]
		\item[$c$] attack vector, $n_b\times 1$;
		\item[$\bar{P},\bar{P}_k$] vectors of monitored line cyber power flows in base case and under contingency $k$, respectively;
		\item[$P_{l,k_t}$] physical power flow on target line $l$ under target contingency $k_t$;
		\item[$P_{G}$] power output of generators, $n_g\times 1$;
		\item[$R_G$] spinning reserve of the generators, $n_g\times 1$;
	\end{description}
	and the parameters are:
	\begin{description}[leftmargin=1.8cm,style=multiline]
		\item[$\sigma$] penalty of the $l_1$-norm of attack vector $c$;
		\item[$G_{B}$] generators to buses connectivity matrix, $n_{b}\times n_{g}$;
		\item[$\textnormal{OTDF}_k$] outage transfer distribution factor matrix under contingency $k$;
		\item[$\textnormal{OTDF}^l_k$] $l^{th}$ row of $\textnormal{OTDF}_k$;
		\item[$N_{1}$] attack vector $l_{1}$-norm limit;
		\item[$L_{S}$] load shift factor, in percentage;
		\item[$H$] dependency matrix between power injection
		measurements and states, $n_{b}\times n_{b}$;
		\item[$P_{D}$] vector of real loads, $n_{b}\times1$;
		\item[$C_{G}$] generation cost vector, $n_g\times 1$;
		\item[$C_{R}$] reserve cost vector, $n_g\times 1$;
		\item[$P_0,P_{k0}$] vectors of pre-SCED monitored line power flows in base case and under contingency $k$, respectively;
		\item[$P_{G0}$] pre-SCED generator outputs, $n_g\times 1$;
		\item[$\textnormal{PTDF}$] power transfer distribution factor matrix;
		\item[$\textnormal{PTDF}^k$] $k^{th}$ row of $\textnormal{PTDF}$;
		\item[$\textnormal{LODF}_k$] line outage distribution factors of monitored lines under contingency $k$;
		\item[$P_{\max}$] vector of base case line limits;
		\item[$P_{k,\max}$] vector of line limits under contingency $k$;
		\item[$P_{G,{\min}}$]  generation lower limits vector, $n_g\times 1$;
		\item[$P_{G,{\max}}$]  generation upper limits vector, $n_g\times 1$;
		\item[$M_G$] ramp rates of all generators, $n_g\times 1$;
		\item[$T_h$] look-ahead time for one period SCED;
		\item[$T_r$] time for spinning reserve requirement.
	\end{description}
	
	Expression in \eqref{eq:Obj1_MaxPF} captures the attacker's objective of maximizing the power flow on line $l$ under target contingency $k_t$, and the penalty factor $\sigma$ is a small positive number to limit the attack size; constraint \eqref{eq:Physical_PF} is the calculation of the power flow on line $l$ under target contingency $k_t$; \eqref{eq:con_L1_norm} models the attacker's limited resources. Ideally, $l_0$-norm should be used to precisely capture the sparsity of $c$, but for tractability reasons we use the $l_1$-norm as a proxy. Constraint \eqref{eq:con_loadshift} limits the percentage of load changes at each bus to avoid detection. 
	
	SCED \eqref{eq:OBJ_MINCOST}-\eqref{eq:con_genctgc} models the system response to the attack. The objective of the operator \eqref{eq:OBJ_MINCOST} is to minimize the total cost, consisting of generation cost and reserve cost; constraint \eqref{eq:con_nodebalance} is the power balance equation; \eqref{eq:con_basePF} is the cyber power flow of the base case monitored lines. Note that this constraint is only modeled for the lines whose pre-SCED power flow is greater than the threshold $\tau$, \textit{i.e., } $|P_0/P_{\max}|\ge \tau$. This is under the assumption that the line flows will not change dramatically after the SCED re-dispatch, due to the ramping constraints of the generators. Similarly, \eqref{eq:con_ctgcPF} is the cyber power flows on monitored lines under each contingency $k$, where $|P_{k0}/P_{k,\max}|\ge \tau$. Here we assume the base case and contingency case monitoring thresholds are the same. In the right hand side of \eqref{eq:con_ctgcPF}, the first term is the pre-SCED post-contingency flows; the second term is the change of the flows as a result of re-dispatch and false loads; the third term represents the amount of power on the monitored lines resulting from the effect of false loads on the contingency line $k$, which is not considered in $P_{k0}$. Constraints \eqref{eq:con_basePmax} and \eqref{eq:con_ctgcPmax} are the line limits in base case and contingency case, respectively. The active power limits in both base case and contingency cases, $P_{\max}$ and $P_{k,{\max}}$, are approximated from the MVA ratings and reactive flows on the branches by
	\begin{flalign}
		P_{\max} = \sqrt{S_{\max}^2 - [\max(Q_{\text{from}}, Q_{\text{to}})]^2}\label{eq:Pmax}\\
		P_{k,\max} = \sqrt{S_{k,\max}^2 - [\max(Q_{k,\text{from}}, Q_{k,\text{to}})]^2}\label{eq:Pkmax}
	\end{flalign}
	where $S_{\max}$ and $S_{k,\max}$ are branch long-term and short-term ratings, respectively; $Q_{\text{from}}$ and $Q_{\text{to}}$ are the base case reactive branch flows at the "from" end and "to" end, respectively; $Q_{k, \text{from}}$ and $Q_{k, \text{to}}$ are those flows in contingency cases.
	Constraints \eqref{eq:con_PGmin} and \eqref{eq:con_PGmax} are the ramping limits; \eqref{eq:con_reserve} is the reserve limit; \eqref{eq:con_genlimit} is the generation limit. Though the RTCA does not simulate generator contingencies, in SCED it is required that when a generator is out, the reserves of all other generators are sufficient to cover the output of the lost generator. We assume the SCED does not include a load shedding policy.
	
	\vspace{-0.2cm}
	\section{\label{sec:Benders}Modified Benders' Decomposition Algorithm to Solve ADBLPs}
	ADBLPs with different objectives in the two levels are in general non-convex. The authors of \cite{Liang2015} solve their ADBLP by replacing the second level defender's problem by its KKT conditions and then convert the problem into an MILP, but this approach does not apply to large-scale systems due to the numerical difficulty brought on by large number of binary variables. To the best of our knowledge, there are no existing techniques to solve large-scale ADBLPs efficiently. In this section, we introduce a modified Benders' decomposition (MBD) algorithm to solve ADBLPs. Benders' decomposition \cite{Benders1962} is an iterative approach to solve linear programs in a distributed manner \cite{ConejoBook}. It is a popular technique to solve optimization problems of large size or with complicating variables. It is also effective in solving complex optimization problems such as stochastic programs and mixed-integer linear programs. In Benders' decomposition, an optimization problem is decomposed into two sub-problems, wherein variables of each sub-problem are treated as constant in the other. The two sub-problems are solved iteratively until the solution converges. Our MBD algorithm modifies the classic Benders' decomposition algorithm to apply it on any ADBLP.
	
	An ADBLP takes the following form (dual variable of the defender's problem is in parentheses):
	\begin{subequations}\label{general}
		\begin{flalign}
			\underset{u}{\text{minimize}} \hspace{0.17cm} & c_1^Tu+d_1^Tv^* \label{generalLv1Obj}\\
		\notag \text{subject to} & \hspace{0.08cm}\\
		& A_1u \ge b_1 \label{generalLv1Con}\\
		& v^*=\text{arg}\{\underset{v}{\text{min}} \hspace{0.17cm} d_2^Tv\} \label{generalLv2Obj}\\
		\notag &\text{subject to} \hspace{0.12cm} \hspace{0.08cm}\\	  
		& \hspace{1cm}A_2u+A_3v \ge b_2  \hspace{0.97cm} (\beta) \label{generalLv2Con}
		\end{flalign}
	\end{subequations}
	where $u$ and $v$ are the attacker's and defender's decision variables, respectively. The defender has no control on $u$, and hence, $u$ in \eqref{generalLv2Con} is treated as a constant in the defender's problem. The attacker does not directly control $v$, but it controls $v^*$ by changing $u$, assuming it has knowledge of the defender's objective and constraints.

	The attack optimization ADBLP \eqref{eq:ADBLP} fits in the form of \eqref{general} where the attack vector $c$ is represented by $u$ and SCED variables $P_G, R_G, P$, and $P_k$ are represented by $v$. In the attacker's objective function, $c_1^Tu$ represents the term $-\sigma \left\Vert c \right\Vert_{1}$, and $d_1^Tv^*$ represents the term $P_{l,k_t}$ in \eqref{eq:Obj1_MaxPF}. Equality constraints can be equivalently written as two inequality constraints. For example, \eqref{eq:con_nodebalance} can be written as
			\vspace{0cm}
		\begin{subequations}
			\begin{flalign}
			\textbf{1}^TP_G &\geq \textbf{1}^TP_D\\
			-\textbf{1}^TP_G &\geq -\textbf{1}^TP_D
			\end{flalign}
		\end{subequations}
	    which fits the form of \eqref{generalLv2Con}. One can similarly map all the constraints in \eqref{eq:ADBLP} to those in \eqref{general}.
	
	The defender's problem \eqref{generalLv2Obj}--\eqref{generalLv2Con}, which represents the system response (SCED) to a fixed attack vector, has the following dual problem (note that $u$ is treated as constant here since it is the fixed attack vector from the attacker's problem):
	\vspace{-0.3cm}
    \begin{subequations}
	\begin{flalign}
		\hspace{0.9cm}\underset{\beta}{\text{maximize}} \hspace{0.17cm} & \beta^T(b_2-A_2u)\label{dualObj2ndSP}\\
		\text{subject to} \hspace{0.12cm}& A_3^T\beta=d_2\label{dualCon12ndSP}\\
		& \beta \ge 0.\label{dualCon22ndSP}
	\end{flalign}
	\end{subequations}
	By weak duality \cite{BoydBook}, for any feasible primal/dual pair, the dual objective value is always less than the primal one: 
	\begin{flalign}
		\beta^T(b_2-A_2u) \le d_2^Tv.	\label{dualnature}
	\end{flalign}
	Since the defender's problem is a linear program, it satisfies strong duality. That is, any feasible point $(v,\beta)$ that satisfies
	\begin{flalign}
		\beta^T(b_2-A_2u) \ge d_2^Tv	\label{PlessD}
	\end{flalign}
	is an optimal solution to it. Therefore, constraints \eqref{generalLv2Con}, \eqref{dualCon12ndSP}, \eqref{dualCon22ndSP}, and \eqref{PlessD} guarantee the optimality of the defender's problem, and hence, can be used to convert the ADBLP to a single level problem as:
	\begin{subequations}
	\begin{flalign}
	\hspace{-0.1cm}\underset{u,v,\beta}{\text{minimize}} \hspace{0.17cm} & c_1^Tu+d_1^Tv \label{Bendersobj}\\
	\hspace{-0.1cm} \text{subject to} \hspace{0.08cm} & A_1u \ge b_1 \label{Benderscon1}\\
	& A_2u+A_3v \ge b_2 \label{Benderscon3}\\
	& A_3^T\beta=d_2 \label{Benderscon4}\\
	& \beta^Tb_2-\beta^TA_2u- d_2^Tv \ge 0\label{Benderscon5}\\
	& \beta \ge 0.\label{Benderscon6}
	\end{flalign}
    \end{subequations}
	The bilinear term $\beta^TA_2u$ in \eqref{Benderscon5} is non-convex and hard to solve. To overcome this difficulty, Benders' decomposition is utilized to decompose this optimization problem into two problems, with $u$ as the variable for the master problem (MP) and $v,\beta$ as the variables for the slave problem (SP). 
	The MP takes the following form:
	\begin{subequations}
		\begin{flalign}
		\hspace{-0.25cm}\underset{u,\alpha}{\text{minimize}} \hspace{0.17cm} & c_1^Tu+\alpha \label{MPObj}\\
        \text{subject to} \hspace{0.08cm}
		& A_1u \ge b_1 \label{MPcon1}
		\end{flalign}
	\end{subequations}
	where $\alpha$ is a variable introduced to represent $d_1^Tv^*$, which will then be updated by adding cuts. 
	The SP is given by:
	\begin{subequations}\label{SP}
	\begin{flalign}
		\underset{v,\beta}{\text{minimize}} \hspace{0.18cm} & d_1^Tv\label{SPObj}\\
        \text{subject to} \hspace{0.11cm}  
		& \beta^Tb_2-d_2^Tv-\beta^TA_2u \ge 0\hspace{0.35cm} (\delta)\\
		& A_3v \ge b_2-A_2u \hspace{1.83cm} (\gamma)\\
		& A_3^T\beta=d_2 \hspace{2.78cm} (\lambda)\\
		& \beta \ge 0.
	\end{flalign}
    \end{subequations}
	At the optimal solution of the SP given by \eqref{SP}, we have
	\begin{flalign}
		d_1^Tv^*=\gamma^Tb_2+\lambda^Td_2-\gamma^TA_2u.\label{SP2Optimal}
	\end{flalign}
	An optimality cut can be added to the MP by taking the right hand side of \eqref{SP2Optimal}:
	\begin{flalign}
		\alpha \ge \gamma^Tb_2+\lambda^Td_2-\gamma^TA_2u. \label{OptCut}
	\end{flalign}
	Note that \eqref{OptCut} is in the MP, and therefore, $u$ is again a variable. If the SP is infeasible with a given $u$, slack variables  $s_i$, $i=1,2,3$, can be introduced to all of the SP constraint to solve the relaxed SP:
	\begin{subequations}\label{SP_relax}
		\begin{flalign}
		\underset{v,\beta, s_i}{\text{minimize}} \hspace{0.18cm} & d_1^Tv\label{SP_relaxObj}\\
		\text{subject to} \hspace{0.11cm}  
		& \beta^Tb_2-d_2^Tv-\beta^TA_2u +s_1\ge 0\hspace{0.35cm} (\hat{\delta})\\
		& A_3v +s_2\ge b_2-A_2u \hspace{1.83cm} (\hat{\gamma})\\
		& A_3^T\beta+s_3=d_2 \hspace{2.76cm} (\hat{\lambda})\\
		& \beta \ge 0.
		\end{flalign}
	\end{subequations}
	where $s_i$, $i=1,2,3$ are the slack variables introduced to ensure feasibility of the relaxed SP. Then, instead of an optimality cut \eqref{OptCut}, a feasibility cut is added to the MP:
	\begin{flalign}
		0 \ge \hat{\gamma}^Tb_2+\hat{\lambda}^Td_2-\hat{\gamma}^TA_2u. \label{FeaCut}
	\end{flalign}
	The MP and SP can then be solved iteratively, with the MP updating $u$ and the SP updating cuts in each iteration.
	\begin{algorithm}[tbh]
		\protect\caption{Modified Benders' Decomposition for Bi-level Linear Programs (MBD)}	\label{alg:Method4}			
		\begin{enumerate}
			\item Set the iteration number $j=1$ and let $u^{(0)}=0$.
			\item Solve the SP \eqref{SP} with $u=u^{(j-1)}$.
			\item If the SP is infeasible, solve the relaxed SP \eqref{SP_relax} and obtain $(\hat{\gamma}^{(j)},\hat{\lambda}^{(j)})$, then add a feasibility cut of form \eqref{FeaCut} to the MP. Otherwise, solve SP \eqref{SP} to get $(v^{(j)},\beta^{(j)},\gamma^{(j)},\lambda^{(j)})$, and add an optimality cut of form \eqref{OptCut} to the MP.
			\item Solve the MP with added cuts and obtain the solution $(u^{(j)},\alpha^{(j)})$.
			\item If $|\frac{d_1^Tv^{(j)}-\alpha^{(j)}}{\alpha^{(j)}}|<\epsilon$, stop. The optimal objective value is obtained as $c_1^Tu^{(j)}+d_1^Tv^{(j)}$. Otherwise, let $j=j+1$ and go to step 2).
		\end{enumerate}
	\end{algorithm}
	
    Solving the SP is equivalent to solving the second level SCED under attack \eqref{eq:OBJ_MINCOST}$-$\eqref{eq:con_genctgc}, while the dual variables of the SP provide information on the objective function \eqref{eq:Obj1_MaxPF}. Since each cut is formulated linearly on the $u$ domain, adding cuts to the MP does not affect its convexity. Thus, MBD is guaranteed to converge in a finite number of iterations \cite{Geoffrion1972}. However, due to the non-convexity of the original bi-level optimization problem, global optimal solution cannot be guaranteed \cite{Sahinidis1991}. Therefore, the optimal objective value obtained by MBD, $\hat{P}_{l,k_t}^{*}$, is a lower bound on $P_{l,k_t}^*$, the global optimal objective. 
	
	\section{\label{sec:Simulation}Simulation Results and Discussions}
	In this section, we present physical consequences through simulations of the attacks designed using the ADBLP described in Sec. \ref{sec:ADBLP}. We use the synthetic Texas system with 2000 buses, 3210 branches, and 432 generators \cite{TexasSystem}. The inputs to the ADBLP described in Sec. \ref{sec:ADBLP} are obtained from OpenPA \cite{OpenPA}, a Java-based EMS simulation platform that we developed in collaboration with our industry partners IncSys \cite{IncSys} and PowerData \cite{PowerData}. Without attack, the system is operating at steady-state, which means that SCED does not change the generation dispatch between each EMS loop. In the base case power flow solution, the total losses among the system is 2\% of the net load. We assume the SCED handles losses by uniformly increasing all loads by this percentage. RTCA simulates contingencies of all branches whose end bus voltages are both at least 100 kV, except radial branches. Prior to attack, RTCA reports no base case warnings nor violations, and 25 post-contingency warnings. We exhaustively design attacks targeting each of those 25 contingency case warnings and test the attack consequences. In our simulations, the short-term branch limit is assumed to be 115\% of the long-term limit, \textit{i.e.,} $S_{k,\max}=115\%\times S_{\max}$; the warning threshold $\tau=90\%$; MBD convergence tolerance $\epsilon=5\times 10^{-5}$; SCED look ahead time $T_h=15$ minutes; spinning reserve time $T_r=10$ minutes. The ADBLP is solved using Matlab with solver CPLEX on a 3.4 GHz PC with 32 GB RAM.
	\vspace{-0.25cm}
	\subsection{Approach for Attack Implementation and System Vulnerability Assessment \label{sec:atk_strategy}}
\begin{figure}[h]
		\centering{}\includegraphics[trim=0 0.2cm 0 0.2cm, scale=0.6]{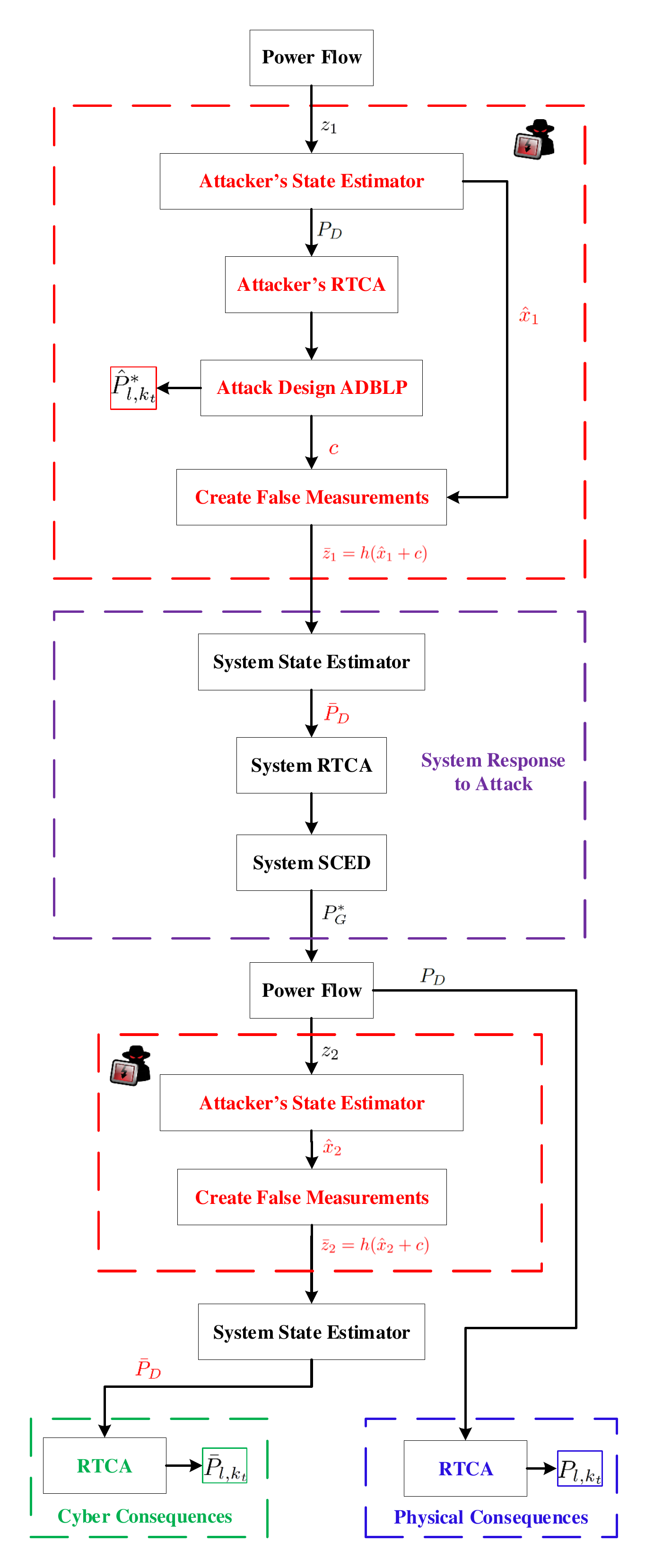}\protect\protect\caption{Attack implementation and system vulnerability assessment approach. \label{fig:atk_strategy}}
		\vspace{-0.6cm}
	\end{figure}

	Fig. \ref{fig:atk_strategy} illustrates the implementation of the attack and the vulnerability assessment approach. For simplicity, we assume that the real loads remain unchanged during the attack period. The physical system behavior and the SCADA measurement collection are simulated by solving an AC power flow. The true measurements $z_1$ from the power flow solution are acquired by the attacker to estimate the states (denoted $\hat{x}_1$). It then performs AC power flow-based RTCA to achieve the security constraints and solves the attack design ADBLP to find the attack vector $c$. Recall that the second level of the ADBLP is a SCED in response to the attack, and by solving it the attacker obtains an estimate on the maximal physical power flow on the target line, which is the optimal objective $\hat{P}_{l,k_t}^{*}$. To implement the designed attack, the attacker then constructs false measurements $\bar{z}_1=h(\hat{x}_1+c)$ and injects $\bar{z}_1$ to the system SE instead of the true measurements $z_1$. Again, only the measurements in the attack subgraph $\mathcal{S}$ are changed. Since the generator outputs are known to the system, the false measurements will cause the SE to estimate a set of false loads. RTCA and SCED are then performed by the system to determine the new optimal generation dispatch $P_G^*$ in response to the false loads. Once the generators re-dispatch, the attacker again acquires the true measurements $z_2$, and estimates the new states $\hat{x}_2$. It then sends $\bar{z}_2=h(\hat{x}_2+c)$ to the system SE to estimate new false loads. The system operator again runs RTCA with the new false loads and observes the cyber power flow $\bar{P}_{l,k_t}$. However, the new dispatch applied on the physical system, will maximize the physical power flow on target line $l$ under target contingency $k_t$, and possibly cause overflow. The true physical power flow, $P_{l,k_t}$, is obtained by running RTCA with the new dispatch and real loads. 
	
	\vspace{-0.35cm}
	\subsection{Results on Maximal Physical Power Flows\label{sec:results_PF}}
\begin{figure}[h]
		\centering{}\includegraphics[trim=0 0.3cm 0 0.3cm, scale=0.48]{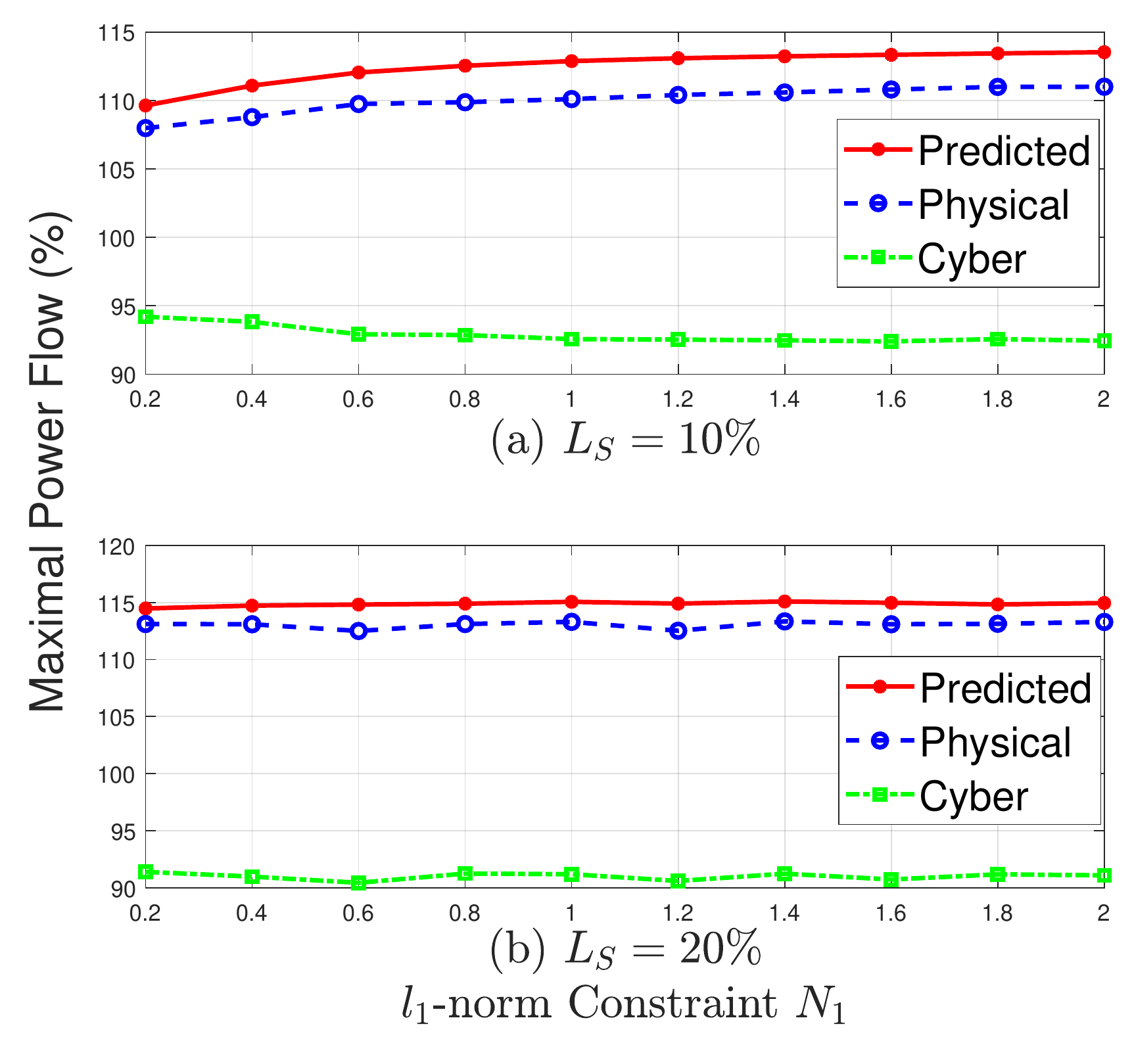}\protect\protect\caption{Comparison of attacker predicted, physical, and cyber power flows on line `ln-2025-2055' under contingency `ln-2054-5236', (a) $L_S=10\%$; (b) $L_S=20\%$ . \label{fig:max_PF}}
		\vspace{-0.4cm}
	\end{figure}

	Fig. \ref{fig:max_PF} compares physical power flow $\hat{P}_{l,k_t}^{*}$ predicted by the attacker, the true power flow $P_{l,k_t}$ in the physical system, as well as the power flow (cyber) seen by the system operator $\bar{P}_{l,k_t}$, as a function of the $l_1$-norm constraint $N_1$. These power flows are plotted as percentage values relative to the active power limit $P_{l,k,\max}$ calculated using \eqref{eq:Pkmax}. The attacker's goal is to maximize the power flow on line `ln-2025-2055' when line `ln-2054-5236' is out of service. 
	
When the load shift $L_S=10\%$, $\hat{P}_{l,k_t}^{*}$ and $P_{l,k_t}$ increase as $N_1$ increases. This indicates that the attacks are effective: they successfully cause post-contingency overflows that cannot be seen by the system operators.
When $L_S=20\%$, similar results are observed, but $\hat{P}_{l,k_t}^{*}$ and $P_{l,k_t}$ are not monotonically increasing as $N_1$ increases. This suggests that the MBD algorithm provides sub-optimal solutions, because as $N_1$ increases, the constraints are relaxed, and the optimal solution for a larger $N_1$ should be at least that of a smaller $N_1$. Maximal power flow is higher when a larger load shift is allowed. With $L_S=20\%, N_1=0.2$, the power flow is higher than that when $L_S=10\%, N_1=2$, which indicates that  in this case load shift is the dominant constraint.

The true physical power flow $P_{l,k_t}$ is slightly lower than the attacker predicted physical power flow $\hat{P}_{l,k_t}^{*}$. One possible reason for this phenomenon is that the attacker is solving a DC approximation of an AC system, and the reactive power flow may change after attack. This could result in a difference in $P_{l,k,\max}$ before and after attack. Another possible reason is that the false measurements $\bar{z}_1$ injected by the attacker cause a different set of security constraints than those that the attacker used to solve the attack design ADBLP. The attacker generates the security constraints by running RTCA using the true measurements, but those constraints generated by the system RTCA are based on the false measurements after attack. As a result, the system SCED solution may be different than the attacker predicted re-dispatch. One approach for the attacker to prevent this situation is to run its own RTCA using the false measurements and include any newly appeared security constraints into the attack design ADBLP, until there are no more new security constraints. However, this approach has no convergence guarantee, and could be too time-consuming to launch the attack in real-time.

Note that in order for the attacks to actually cause post-contingency violations requires a particular contingency to occur. Thus, the attacker has to create the target contingency itself, or gain insider knowledge about when the contingency is likely to occur. Both are plausible for sophisticated attackers. More aggressively, the attacker can aim to create base case overflows, but the $N-1$ reliable constraints may push the system to operate conservatively. In the synthetic Texas system, there is no branch whose base case power flow is higher than $\tau$ prior to the attack. Thus, to cause base case overflow, the attacker has to shift a tremendous amount of load that may easily trigger an alarm at the control center. Moreover, a large load shift will move the system operating condition dramatically with high probability, and thereby create new security constraints that are not considered when designing the attack. Thus, the consequences of the attack become unpredictable for the attacker. We have attempted to design a base case attack targeting branch `ln-7058-7095' that has the highest base case power flow in percentage, but no overflow can be found with $L_S=90\%$ and $N_1=20$. With $L_S=100\%$ and $N_1=20$, the attacker's predicted power flow reaches $102.29\%$, but the false measurements create 3197 warnings and 24773 violations at the RTCA solution.

\vspace{-0.2cm}
	\subsection{\label{sec:L0result}Results on Attack Resources}
\begin{figure}[h]
		\centering{}\includegraphics[trim=0 0.3cm 0 0.3cm, scale=0.45]{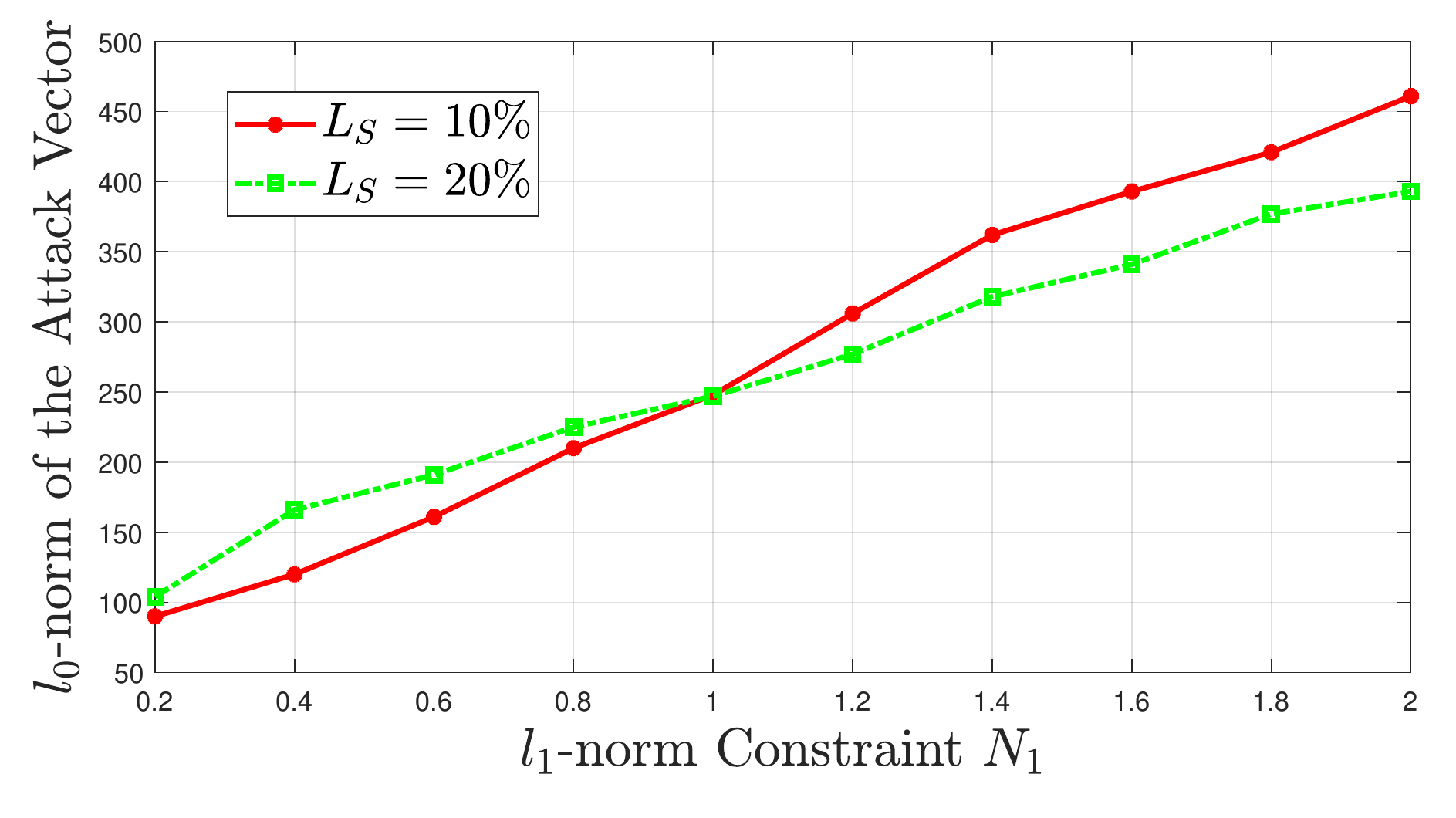}\protect\protect\caption{Comparison of the $l_0$-norm of the attack vector for target line `ln-2025-2055' under contingency `ln-2054-5236'. \label{fig:L0LS}}
		\vspace{-0.8cm}
	\end{figure}

	Fig. \ref{fig:L0LS} illustrates the relationship between maximal power flow and $l_0$-norm of the attack vector (\emph{i.e.} the number of center buses in the attack) versus the $l_1$-norm constraint $N_1$ for target line `ln-2025-2055' under contingency `ln-2054-5236', with different load shift constraints. As $N_1$ increases, so does the $l_0$-norm of the attack, indicating that $l_1$-norm is a valid proxy for $l_0$-norm for our problem. If a larger load shift is allowed, the maximal power flow on target line increases, but the resulting $l_0$-norm may decrease for the same $N_1$. This indicates a trade-off between load shift and attacker's resources: as the attacker attempts to avoid detection by minimizing load changes, it will require control over a larger portion of the system to launch a comparable attack.
	\vspace{-0cm}
	\subsection{Comparison of Physical and Cyber RTCA results\label{sec:scatter}}
\begin{figure}[h]
		\centering{}\includegraphics[trim=0 0.3cm 0 0.3cm, scale=0.35]{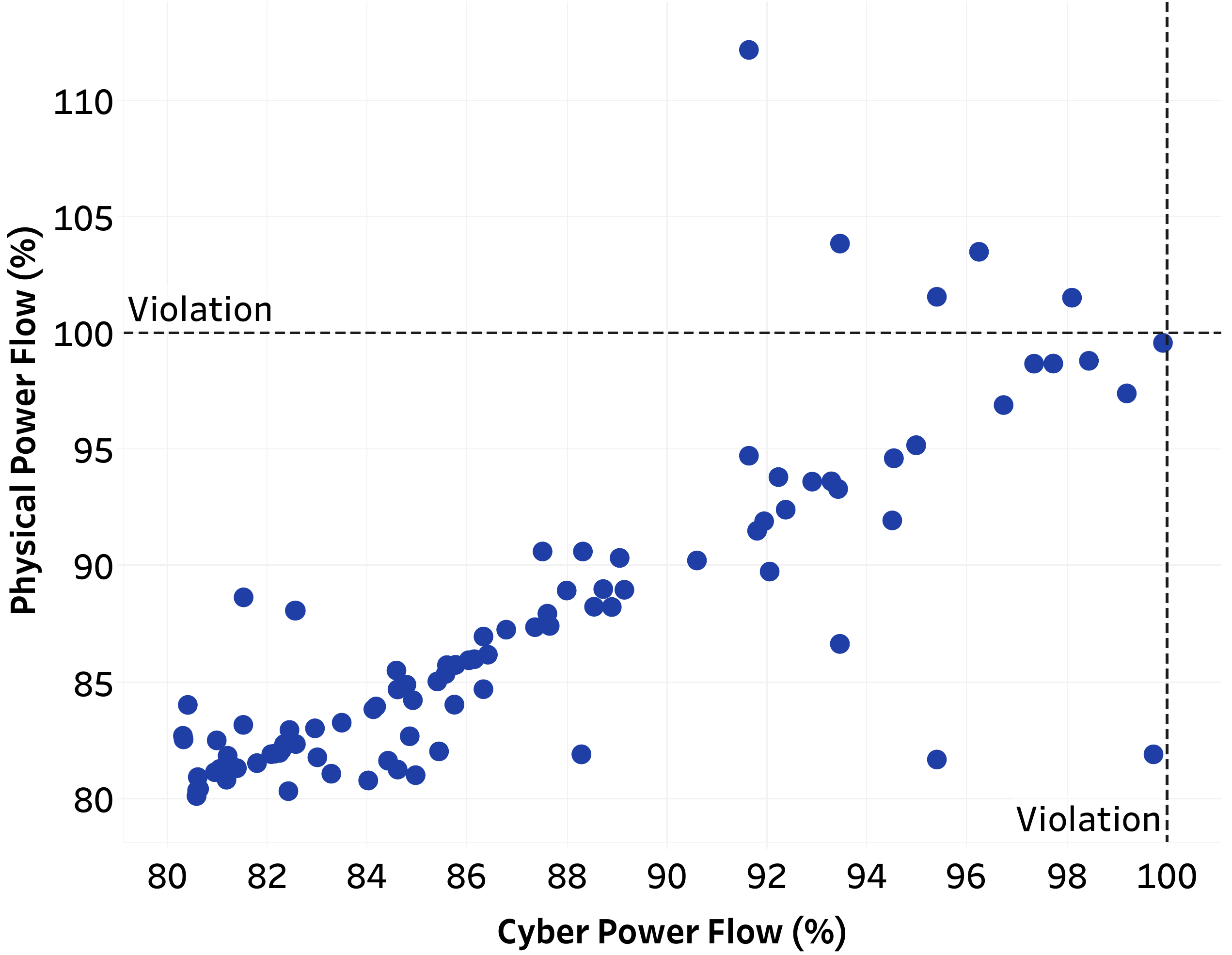}\protect\protect\caption{Comparison of the physical and cyber RTCA results after re-dispatch. \label{fig:scatter}}
	\vspace{-0.6cm}
	\end{figure}

	Fig. \ref{fig:scatter} compares the physical and cyber RTCA results after the re-dispatch resulting from an attack on target line `ln-2025-2055' under contingency `ln-2054-5236' with load shift $L_S=10\%, N_1=2$. The cyber post-contingency power flows on the x-axis represent what the system operator observes, while the y-axis represents the post-contingency power flows in the physical system. There is no point beyond 100\% of the x-axis, which indicates that the system operator sees no post-contingency violation after the attack. Therefore, the attack successfully spoofed the operator that the system is in a secure state, while in reality, the target line has a 112.2\% post-contingency overflow. In addition, there are four post-contingency violations that are caused by the same attack, even though they are not the attacker's targets.
	
	\subsection{Statistical Results on Attack Consequences\label{sec:result_stat}}
	As mentioned at the beginning of Sec. \ref{sec:Simulation}, we exhaustively tested attacks targeting the 25 branches with post-contingency warnings. The designed attacks successfully cause overflows on 8 out of the 25 target branches. Table \ref{tab:result_stat} gives the statistical results on attack consequences of these 8 branches. We derived attacks using $l_1$-norm constraints in the range from $N_1=0.2$ to $N_1=2$. The table shows the resulting ranges in maximal power flow and $l_0$-norm of the attack vector $c$ across this range. The load shift constraint $L_S=10\%$. The prefix `ln' indicates a transmission line and `tx' indicates a transformer. From the maximal power flow range, we can see that some branches are more vulnerable than others, as they have higher overflows. Thus, the system operators can identify critical lines and critical contingencies for attack protection purposes. For example, they can artificially
	reduce the line limit to keep the attack from being successful.
	Measurements around vulnerable branches can be encrypted to
	prevent them from being modified. In our ADBLP,
	the load shift constraint characterizes the detectability of
	the attack, indicating that load abnormally detectors can help
	system operators distinguish between natural load changes and
	possible cyber attacks based on load redistribution.
	\begin{table}[h]
		\renewcommand{\arraystretch}{1.3}
		\protect\caption{Statistical Results on Maximal Physical Power Flow and $l_0$-norm of the Attack Vector with $N_1\in [0.2,2]$ \label{tab:result_stat}}
		\centering
		\begin{tabular}{|c|c|c|c|}
			\thickhline
			Target & Contingency & Max PF Range (\%) & $\|c\|_0$ Range \tabularnewline
			\hline
			ln-6188-7305 & ln-7058-7095 & 101.92--105.08 & 133--442 \tabularnewline
			\hline
			ln-6240-6287 & ln-6141-6239 & 102.43--106.76 & 137--314 \tabularnewline
			\hline
			ln-7233-7251 & tx-6063-6062 & 105.41--107.90 & 156--485 \tabularnewline
			\hline
			ln-1003-1055 & ln-3046-3078 & 102.80--102.94 & 163--520 \tabularnewline
			\hline
			ln-2025-2055 & ln-2054-5236 & 107.98--111.00 & 90--461 \tabularnewline
			\hline
			ln-2070-5237 & ln-2054-5236 & 101.35--104.35 & 90--461 \tabularnewline
			\hline
			ln-1003-1055 & ln-1004-3133 & 102.43--102.56 & 160--513 \tabularnewline
			\hline
			ln-7059-7407 & ln-7058-7406 & 100.38--102.24 & 154--488 \tabularnewline
			\thickhline 
		\end{tabular}
		\vspace{-0.5cm}
	\end{table}



	\section{\label{sec:conclusion}Conclusion}
	We have evaluated the vulnerability of $N-1$ reliable power systems to unobservable FDI attacks via the physical consequences of such attacks. Such N-1 reliable systems are assumed to be operated by an EMS consisting of SE, RTCA, and SCED. The attacker injects intelligently designed false measurements to the SE that bypass the bad data detector, and cause the SE to estimate false loads. The SCED re-dispatch resulting from the false loads leads to the power flow on a target line (picked by the attacker) to be maximized.
	
	We have also highlighted the knowledge required by the attacker to design such attacks. In the worst case, the attacker can perform exactly the same RTCA and SCED as the system does, and hence, can approximately predict the system response to the attacks. Designing these attacks involves solving an ADBLP that is non-convex and difficult to solve in general. An efficient algorithm based on Benders' decomposition is introduced to solve the attack design ADBLPs. 
	
	The designed attacks can successfully cause post-contingency overflows on target branches. Moreover, they may create more violations on branches other than the target one. Our vulnerability assessment approach can help system operator identify critical branches and critical contingencies to design protection schemes. Future work will include designing countermeasures to detect, identify, and mitigate such attacks.

	\section*{Acknowledgment}
	This material is based upon work supported by the National Science Foundation under Grant No. CNS-1449080, and grant S-72 from the Power System Engineering Research Center (PSERC). We would like to thank the following at ASU: Mr. Andrea Pinceti for creating the base case, Mr. Roozbeh Khodadadeh for help with the test platform, and Prof. Kory Hedman and his team for their support with RTCA and SCED. We also thank Dr. Robin Podmore (IncSys) and Mr. Christopher Mosier (Powerdata) for the OpenPA software. 
	

	%
	%

	
	
	%
	%
	%
	
	\bibliographystyle{IEEEtran}
	\bibliography{dis}

\end{document}